\documentclass[10pt,conference,a4paper]{IEEEtran}

\usepackage[np]{numprint}
\usepackage[utf8]{inputenc}
\usepackage[T1]{fontenc}
\usepackage[dvipsnames]{xcolor}
\usepackage{graphicx}
\usepackage{arydshln}
\usepackage{booktabs}
\usepackage{caption}
\usepackage[caption=false]{subfig}
\usepackage{arydshln}
\usepackage{multirow}
\usepackage{import}
\usepackage{amssymb}
\usepackage{fp}
\usepackage{comment}
\usepackage{adjustbox}
\usepackage{balance}
\usepackage[breaklinks]{hyperref}
\usepackage[sort,nocompress]{cite}
\usepackage{cleveref}
\usepackage{xspace}
\usepackage[binary-units]{siunitx}
\DeclareSIUnit{\packet}{p}
\usepackage[nolist]{acronym}

\usepackage{todonotes} %
\presetkeys{todonotes}{fancyline, color=olive!30, inline}{}
\newcommand{\budget}[1]{\todo[color=yellow!40]{\textit{Budget: #1 page(s)}}}
\renewcommand{\budget}[1]{}

\usepackage{pifont} %
\usepackage[normalem]{ulem}
\usepackage{microtype}

\usepackage{blindtext}

\usepackage{floatrow}
\newfloatcommand{capbtabbox}{table}[][\FBwidth]

\newcommand{\eg}{e.g.,\xspace}
\newcommand{\ie}{i.e.,\xspace}

\newcommand{\parx}[1]{\noindent\textbf{#1}\xspace}
\newif\ifcutoptional
\cutoptionalfalse
\DeclareSIUnit{\nothing}{\relax}

\newcommand{\roughly}[1]{\char`\~{#1}}

\renewcommand*{\paragraph}[1]{%
    \vspace{.5em}
    \noindent
    {\normalfont \bf #1}
}

\begin{document}

\title{Analyzing the IoT Ecosystem in the IPv6 Internet}
\title{A Multidimensional Analysis of \\IoT Hosts in the IPv6 Internet}
\title{Analyzing IoT Hosts in the IPv6 Internet}

\author{\IEEEauthorblockN{Peter Jose}
\IEEEauthorblockA{Max Planck Institute for Informatics\\
pjose@mpi-inf.mpg.de}
\and
\IEEEauthorblockN{Said Jawad Saidi}
\IEEEauthorblockA{Max Planck Institute for Informatics\\
jsaidi@mpi-inf.mpg.de}
\and
\IEEEauthorblockN{Oliver Gasser}
\IEEEauthorblockA{Max Planck Institute for Informatics\\
oliver.gasser@mpi-inf.mpg.de}}

\maketitle

\begin{abstract}
Users and businesses are increasingly deploying Internet of Things (IoT) devices at home, at work, and in factories.
At the same time, we see an increase in the use of IPv6 for Internet connectivity.
Even though the IoT ecosystem has been the focus of recent studies, there is no comprehensive analysis of IoT end-hosts in the IPv6 Internet to date.

In this paper we perform an in-depth analysis of IPv6-reachable IoT hosts using active measurements.
We run measurements targeting 530M IPv6 addresses on six popular IoT-related protocols.
With 36.4K hosts in 156 countries we find 380$\times$ fewer IoT-speaking end-hosts compared to IPv4.
Moreover, we conduct a security analysis for TLS-enabled IoT-speaking hosts identifying up to 57\% untrusted certificates, with up to 32\% being self-signed and 25\% being expired.
Finally, we plan to publish our measurement results, tools, and a website dashboard to foster further research in the area.
\end{abstract}
\section{Introduction}

The Internet of Things (IoT) has become increasingly prevalent over the last few years.
Consumers use IoT devices for entertainment (\eg smart TVs), home automation (\eg Google Home, Amazon Alexa), or home surveillance (\eg Ring Home Security System).
In addition, the industry is also making use of IoT (\ie industrial IoT) to automate production processes.
Studies estimate that by 2025 there will be 28 billion IoT devices~\cite{IoT-stats}.

Consequently, the IoT ecosystem has been the focus of research in recent years.
Several studies have analyzed various aspects of the IoT ecosystem, such as identifying IoT devices in the Internet~\cite{saidi2020haystack, guoIoT2020, pashamokhtari2021inferring, http-scan-iot,iotfinder }, leakage of privacy-sensitive information by IoT devices~\cite{saidi2022badapple, Information-Exposure-IMC2019}, and servers contacted by IoT devices~\cite{Information-Exposure-IMC2019, Saidi2022}

Despite the growing importance of IPv6~\cite{googleipv6-adoption, ipv6roles}, the IoT ecosystem in the IPv6 Internet has remained understudied.
This may be due to various factors such as difficulty in identifying target addresses in the vast IPv6 address space~\cite{gasser2018clusters, gasser2016ipv6scanning}, a lack of IPv6 support of IoT clients until recently~\cite{iot_ipv6_adoption}, and relatively low deployment of IPv6 in ISP networks~\cite{ipv6adoption}.
Moreover, commercial network intelligence platforms such as Censys only started to release IPv6 data recently~\cite{censys-ipv6}.
Similar to IPv4 ~\cite{Shreyas2021}, however, we find these platforms to have relatively low coverage of IoT-protocol-speaking hosts in the IPv6 Internet.

Due to these shortcomings in the state-of-the-art, we currently lack a good understanding of IoT-speaking end-hosts in IPv6.
Therefore, this paper aims to address this gap by detecting, characterizing, and analyzing the deployment of IoT-speaking end-hosts in the IPv6 Internet using active measurements.
Specifically, this work makes the following main contributions:

\begin{itemize}
    \item \textbf{IoT IPv6 deployment:} We perform a large-scale active measurement study on the IPv6 Internet.
        We target 530$M$ IPv6 addresses for six IoT-related protocols (AMQP, CoAP, MQTT, OPC~UA, XMPP, and Telnet) running on eleven ports.
        In total, we find 36.4$K$ hosts in 156 countries and 3177 ASes, respectively.
        Moreover, Telnet is the most frequent one in our measurements, and we find support for multiple protocols on the same host to be relatively rare.
        Furthermore, we find IoT-speaking end-hosts to be relatively stable in terms of responsiveness 62.5\%.
    \item \textbf{Security analysis:} Next, we perform an in-depth study of TLS security properties of IoT-speaking end-hosts.
        For XMPP and MQTT, we find that most of these end-hosts support the latest version of TLS, whereas this is only the case for less than 40\% of AMQP end-hosts.
        We analyze TLS certificates sent by these end-hosts and find that 19.2\% of them are self-signed and 45.0\% are untrusted.
        Additionally, we find that 25.0\%, 23.7\%, and 13.0\% of certificates are expired for AMQPs, MQTTs, and XMPPs, respectively.
    \item \textbf{Measurement tools and dashboard:} To run our active measurement study, we develop custom extensions to ZMap and ZGrab2.
        To foster further research in the IoT ecosystem by fellow scientists, we publish our custom-developed measurement tools, including ZMap probe packets for DTLS and ZGrab2 modules for AMQP, CoAP, DTLS, MQTT, and XMPP \cite{zgrab2modules}.
        Moreover, we plan to publish raw measurement results to foster reproducibility in the Internet measurement community.
        Finally, we provide a publicly accessible dashboard as an easy way to interact with our results \cite{dashboard}.
\end{itemize}

\section{Background}

\begin{figure*}[!t]
    \centering
    \includegraphics[width=\textwidth]{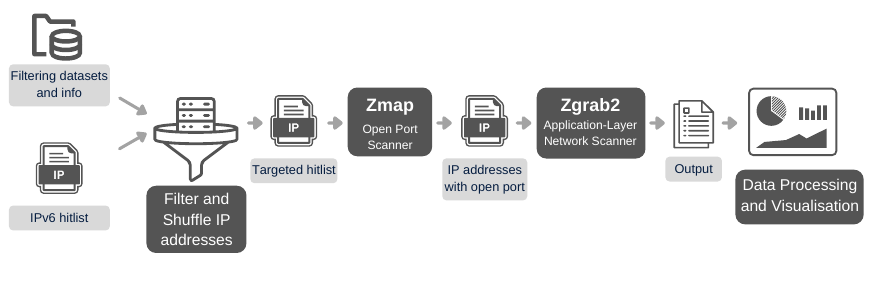}
    \caption{Scanning pipeline for discovering IoT-speaking hosts in the IPv6 Internet.}
    \label{f:pipeline}
\end{figure*}

This section provides background information on the popular IoT protocols considered in this study. Internet and sensing networks combine to form the IoT paradigm allowing machine-to-machine communication~\cite{internet_of_things}. An IoT network consists of IoT devices and servers interacting and performing their respective roles. IoT devices have different computation power ranging from simple embedded sensors with limited resources (computational power, energy, and memory) to advanced and powerful ones. Usually, the protocols developed for IoT devices tend to be lightweight to accommodate constrained resources~\cite{iot_energy,internet_of_things}. 

The most commonly used protocols in IoT applications that we scan are the Message Queuing Telemetry Transport \textbf{(MQTT)} protocol, the Constrained Application Protocol \textbf{(CoAP)}, the Extensible Messaging and Presence Protocol \textbf{(XMPP)}, the Advanced Message Queuing Protocol \textbf{(AMQP)}, the Open Platform Communication Unified Architecture \textbf{(OPC UA)} protocol, and the Telnet protocol~\cite{iot_survey,iot_protocols,iot_protocols_vulnerabilities}. These protocols use both TCP and UDP as their transport protocols. However, most of them primarily use TCP. For CoAP, UDP is recommended for constrained node networks owing to the larger resource requirement for CoAP over TCP~\cite{rfc8323}. Protocols over TCP use TLS, and those over UDP use DTLS in their secured versions.
An exception is Telnet, which does not provide a secure version.

\parx{MQTT:}
MQTT follows a client-server publish/subscribe model designed in a lightweight event-driven approach making it ideal for usage in constrained environments.

\parx{CoAP:}
CoAP follows the client-server interaction model inspired by the Hypertext Transfer Protocol (HTTP). CoAP provides an option to increase the communication's reliability by using a confirmable message.

\parx{XMPP:}
XMPP is an open-source messaging protocol designed originally for text messaging and application-to-application messaging. It is an Extensible Markup Language (XML) and text-based protocol that uses both request/response and publish/subscribe architectures over TCP. Data in the form of XML stanzas~\cite{rfc3920} are used in the communication between the XMPP client and server. 

\parx{AMQP:}
AMQP is a message-oriented, lightweight, and open-source protocol. This protocol is designed for the publish/subscribe and request/response architectures~\cite{iot_protocols_vulnerabilities}.

\parx{OPC UA:}
OPC UA is a cross-platform, open-source communication protocol that aims at the Service Oriented Architecture (SOA). The protocol supports both publish/subscribe and client/server methods. The entire specification of the OPC classic is enhanced and unified by the creation of OPC~UA. The protocol is mainly deployed for industrial applications~\cite{iot_protocols_vulnerabilities}. 

\parx{Telnet:}
Telnet provides an application-layer protocol providing bi-directional text-oriented communication via a terminal service. Due to the simplicity of Telnet, it is still very popular on embedded systems, even though it is not heavily used on servers and workstations anymore~\cite{telnet_paper}.
\section{Methodology}

\label{methodology}
In this section, we detail our methodology on identifying IoT-protocol speaking hosts in the IPv6 Internet. For this, we need to address several challenges namely, the vast IPv6 space address space, IoT protocol selection, and the lack of support for some IoT protocols in existing tools. We start by generating IPv6 targets. Next, we explain the rationale behind selecting the IoT protocols and finally the implementation details of our scanning tools. ~\Cref{f:pipeline} illustrates an overview of our methodology.

\subsection{Target Generation}

The vast address space of IPv6 makes it infeasible to scan all of it. Hence, we use the publicly available hitlist provided by Gasser et al.~\cite{gasser2016ipv6scanning,zirngibl2022rustyclusters}, which is comparatively less biased and provides supplementary data aiding the filtration of aliased prefixes from the hitlist. 

In the filtering step, we first identify and remove addresses in aliased prefixes. This helps us avoid overcounting hosts, i.e., avoiding a single machine responding to all addresses in a prefix. For this, we employ multi-level aliased prefix detection technique~\cite{gasser2018clusters}.    

Next, complying with the best practices in active measurement, we remove prefixes which are listed on our internal blocklist. The blocklist contains prefixes from previous studies, where network administrators have asked us not to scan their addresses. See~\Cref{sec:ethical} for more details on ethical considerations. From here on, we refer to this filtered hitlist as the targeted hitlist.

\subsection{Protocol Selection}

Communication protocols are an integral part of IoT systems. The selection of one protocol suitable for different IoT applications is faced with several dilemmas that need to consider energy efficiency, security, and quality of service. Recent studies dealing with IoT-speaking hosts using active measurement considered protocols such as MQTT, AMQP, UPnP, CoAP, XMPP, and Telnet\cite{iot_survey,iot_protocols, Shreyas2021}.
In our work, except for UPnP, we also consider these protocols.
 Moreover, we consider OPC~UA, a unified version of the OPC classic with its service-oriented architecture \cite{iot_protocols}. For each protocol---with the exception of Telnet---, we consider both the secured and non-secured version. Considering these factors, we select the five protocols listed in \Cref{t:ProtocolPort}. 
Although some protocols can use TCP and UDP as transport layer protocols, we focus on only the most recommended ones. For example, using CoAP over UDP is recommended for constrained node networks owing to the larger resource requirement for CoAP over TCP~\cite{rfc8323}. 

Although IANA has specified standard secured and unsecured ports for most of our protocols, a protocol can still be served on non-standard ports~\cite{LZR2021}. For XMPP, we consider the port corresponding to the client communication. Moreover, IANA has yet to specify the secured port of XMPP; hence we use its conventional port 5223~\cite{rfc3920}. In our study, except for XMPP, we only consider the standard ports for all the protocols. For a complete list of protocols and their ports, we refer to~\Cref{t:ProtocolPort}.

In the subsequent sections, we refer to the secure version of each protocol with a suffix ``s'', \eg the secure version of XMPP is referred to as ``XMPPs''.

\begin{table}[!t]
    \caption{TCP/UDP based categorization of protocols and ports considered for our measurements.}
    \label{t:ProtocolPort}
    \centering
    \begin{tabular}{ccc cc}
        \toprule
        Protocol & Unsecured Port & Secure Port & Transport \\
        \midrule
        CoAP  & 5683 & 5684 & UDP \\ 
        MQTT   & 1883 & 8883 & TCP \\
        XMPP   & 5222 & 5223	& TCP \\ 
        AMQP  & 5672 & 5671	& TCP \\
        OPC~UA  & 4840 & 4843 & TCP \\ 
        Telnet & 23 & n/a	& TCP \\
        \bottomrule
    \end{tabular}
\end{table}

\subsection{Open Port Scans}
\label{Open Port Scan Stage}

Next, we scan for open ports on the targeted hitlist using a modified version of ZMap~\cite{zmappaper} that supports IPv6~\cite{zmapv6}, targeting one IoT port at a time. Among the various options provided by ZMap, we use two scan options: (1) TCP SYN scans to identify ports supporting TCP-based protocols and (2) application-specific UDP scans for ports supporting UDP-based protocols. Further, to support UDP-based protocols, we create custom probes for CoAP and DTLS. We send one probe per protocol per target IP address and record all IP addresses with at least one successful response.

\subsection{Application-Layer Network Scans}
\label{l:me_application_layer_network Scan Stage}

Using the IP addresses with successful responses from the open port scan, we conduct application-layer handshakes for each protocol using ZGrab2~\cite{zgrab}.
Except for Telnet, the default implementation of ZGrab2 does not support any of the protocols listed in Table~\ref{t:ProtocolPort}. We extend ZGrab2 by adding corresponding modules. 

Furthermore, we modify ZGrab2 by adding TLS 1.3 support to analyze all standardized TLS versions.
We also extend ZGrab2 to support DTLS 1.2, by incorporating the library by Pion~\cite{dtls} to add support for a secure version of our UDP-based protocols. We plan to release our  ZGrab2 tools and modules to facilitate further research on IoT protocols and other UDP-based protocols.

In total we run three measurement campaigns over a period of six months, spanning from January to June 2022.

\subsection{Limitations}

Even though we use a multi-step scanning pipeline, our approach has limitations.
First, protocols such as AMQP, XMPP, and Telnet can also be used for other non-IoT applications.  %
 Second, we do not differentiate between IoT devices and servers. Therefore, further investigation is needed to distinguish between them.
Third, our coverage of IPv6 hosts is limited to our underlying hitlist. Finally, unlike the recent work in IPv4~\cite{Shreyas2021}, we do not apply  honeypot detection techniques to infer whether a given target is a honeypot or not. %

\begin{figure}[!t]
    \centering
    \includegraphics[width=\columnwidth]{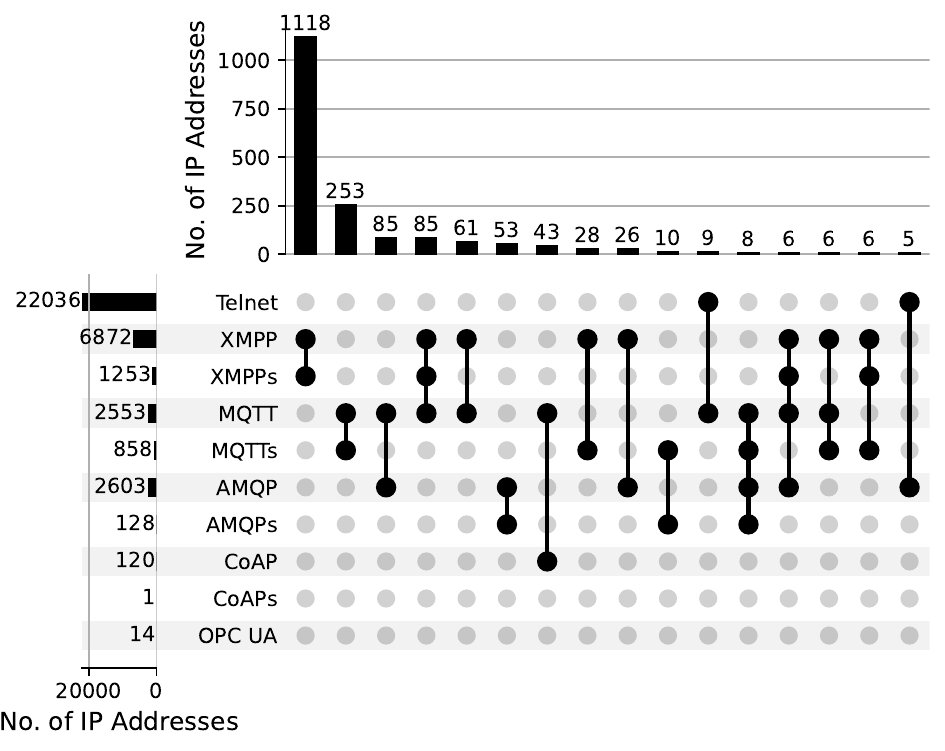}
    \caption{Upset plot with the horizontal section showing the overall number of responsive IP addresses per protocol and the vertical section showing the number of IP addresses hosting different protocol combinations.}
    \label{f:pp2}
\end{figure}

\subsection{Ethical considerations}
\label{sec:ethical}
We take several measures to ensure that our scanning does not cause harm to routers or networks. To minimize the impact, we use a low measurement load, sending only one packet per destination and port. In addition, as by the policy of our institution, we filter out IP addresses belonging to Russia due to the Ukraine-Russian war to avoid raising alarms. We also randomly spread the load at each target IPv6 in the hitlist and coordinate with local network administrators to ensure that our scanning does not harm the local or upstream network.

\begin{figure*}[!t]
    \centering
    \includegraphics[width=\textwidth]{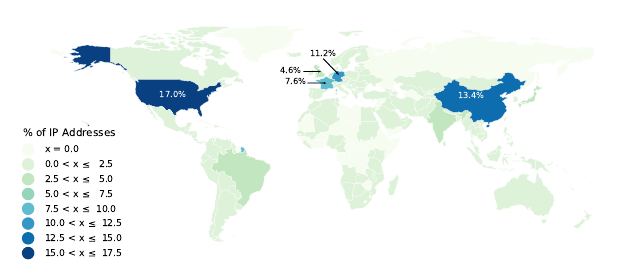}
    \caption{Distribution of IP addresses hosting popular IoT protocols based on geographical location.}
    \label{f:map}
\end{figure*}

To ensure that our active scanning is ethical and does not result in complaints or opt-out requests, we follow the best current practices~\cite{zmappaper,measurementPractices1,measurementPractices2}. One of these practices is ensuring that our prober IP address has a meaningful DNS PTR record. We also set up a web server with experiment and opt-out information on the measurement machine.
During our active experiments we did not receive any complaints or opt-out requests.

\section{IoT Hosts in the IPv6 Internet}

\budget{2}

In this section, we go through the results from our Internet-scale active measurements using different IoT protocols. 
We discuss the characteristics of the responsive hosts from a protocol, geographical, temporal, and network type perspective.

\subsection{Responsive Protocols}

We use results from our latest measurement (2022-06-29) to analyze the responsiveness of different protocols.
\Cref{f:pp2} shows the the popularity of each protocol in terms of the number of hosting IP addresses (horizontal bars) and the number of IP addresses hosting different protocol combinations (vertical bars).
Although we observe 37 combinations of protocols in our data, we only illustrate those with at least two ports and five occurrences for readability.

Consistent with recent work in IPv4~\cite{Shreyas2021}, we observe that the Telnet protocol is the most popular with $\sim$22K IPs. However, we also find a relatively high number of XMPP-speaking hosts, making XMPP the second most popular protocol. We notice that there are particularly fewer CoAP hosts in IPv6 in comparison to previous IPv4 work.
When doing a protocol-by-protocol comparison, we find 380$\times$ fewer IoT-speaking end-hosts in IPv6 compared to IPv4 (34.2K vs. 13M).

\subsection{Geographical Characterization}

Next, we present our analysis of responsive hosts from a geographical perspective based on our latest measurement (2022-06-29).  We geo-locate responsive hosts using the MaxMind GeoLite2 database~\cite{geolite} on a country level. We first analyze the contribution of responsive IPv6 addresses per country, and subsequently, we drill-down and analyze the country-protocol distribution of IPv6 addresses.

In~\Cref{f:map}, we examine the geographical distribution of the IP addresses that host at least one of our studied protocols. Our observation is that the distribution is skewed in terms of the number of countries represented. Specifically, we find that the United States, China, Germany, France, and the United Kingdom to account for more than 53\% of all observed IP addresses. For the remaining countries, we can see a long tail in the distribution, with IP addresses distributed among 151 countries and no single country accounting for more than 4\% of the total IPs.

\begin{figure}[!t]
    \centering
    \includegraphics[width=0.95\columnwidth]{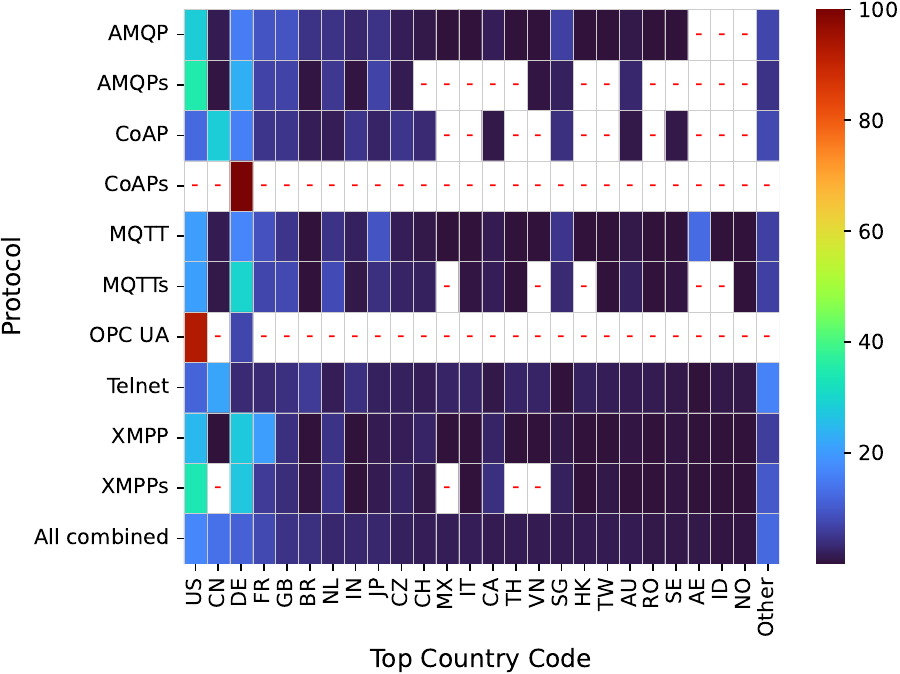}
    \caption{Heatmap distribution of IP addresses hosting popular IoT protocols in different countries.}
    \label{f:heatmap}
\end{figure}

Next, we analyze per-country differences for responsive protocols.
In \Cref{f:heatmap}, we break down the percentage of IP addresses that each country contributes for each protocol. We show the top 25 countries and group the remaining countries into the ``Other'' category. Our first takeaway is that all top 25 countries have some occurrences of our top protocols, namely, XMPP and MQTT. For XMPP, we also observe that the majority of IPs are in the United States. However, we find that for Telnet, 21.5\% of hosts are attributed to China, which is in line with previous studies focusing on IPv4~\cite{iotinsecurity2020}.

\subsection{Temporal Characterization}

Next, we analyze temporal stability of IPv6 IoT-speaking end-hosts.
To investigate this, we repeat our experiments three times over a six-month period, with each experiment conducted roughly two months apart. We first present an overview of the total number of responsive hosts in each of the three scans and then focus on the changes in the number of responsive hosts between subsequent scans.

\begin{figure}[!t]
    \centering
    \includegraphics[width=0.95\columnwidth]{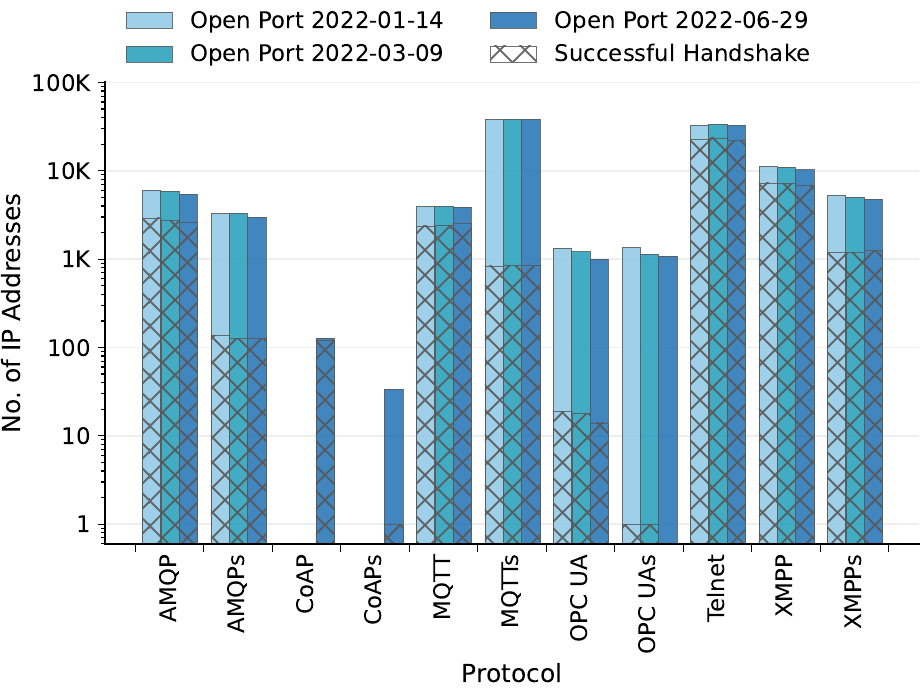}
    \caption{Comparison of open ports and successful handshakes across our three measurements.}
    \label{f:open_ports}
\end{figure}

In \Cref{f:open_ports}, we illustrate the number of hosts discovered at each scan for different protocols. Each bar corresponds to the number of hosts with an open port for the respective protocol on the scan date. We annotate the number of hosts with a successful handshake by applying a lattice-like pattern. In the first two scans, our toolchain did not support the CoAP protocol, hence we only show the result for the latest scan.

From January to July 2022, we observe a decline in the total number of responsive IPv6 addresses in our underlying hitlist \cite{gasser2018clusters}, before increasing again in July 2022. In line with this, we observe a slight decline in the number of responsive hosts for the majority of the protocols. Nevertheless, the proportion of successful handshakes remains consistent. For the MQTT and MQTTs protocols, the number of hosts with an open port remains stable and we also observe a slight increase in the proportion of successful handshakes.

Another notable observation is the substantial difference in the proportion of successful handshakes in secure and non-secure versions of each protocol. We notice a higher rate of successful handshake in non-secure versions compared to the secure versions. For example, in AMQP, we perform a successful protocol handshake for 48.3\% of hosts with open ports, while in its secure version, the success rate is only 4.3\%. Upon investigation, we track down these differences to TLS handshake fails, as in the majority of cases we do not reach the actual protocol handshake. The explanations for such behaviors can be that end-hosts may use Server Name Indication (SNI)~\cite{rfc6066} or expect their clients to present certificates to complete the TLS handshake~\cite{Saidi2022}.

\begin{figure}[!t]
    \centering
    \includegraphics[width=0.95\columnwidth]{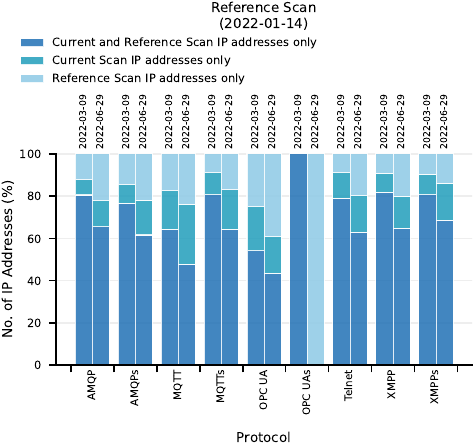}
    \caption{Churn of IP addresses for consecutive scans with the first scan used as a reference.}
    \label{f:s3}
    \vspace{-0.5cm}

\end{figure}

Next, we analyze the address churn, \ie the change in responsive IPv6 addresses between our measurements.
In~\Cref{f:s3}, we present the percentage change in the number of responsive hosts relative to the first scan (reference scan). The bars represent subsequent scans and indicate the percentage of IP addresses that are observed in both scans, the reference scan only, and the subsequent scan only. Our observations show that for all protocols, the latest scan on June 29, 2022, has the smallest overlap with the reference scan. This is expected since the two scans are conducted almost six months apart. Nevertheless, there is still an overlap of at least 40\% for most protocols. One exception is the secure OPC UA protocol, where we see no overlap in the latest scan, since the single responsive host in the first two scans did not respond anymore.
Overall, when we consider all protocols, we observe an overlap of 62.5\% between the first and last scan.

\begin{figure}[t]
    \centering
    \includegraphics[width=0.95\columnwidth]{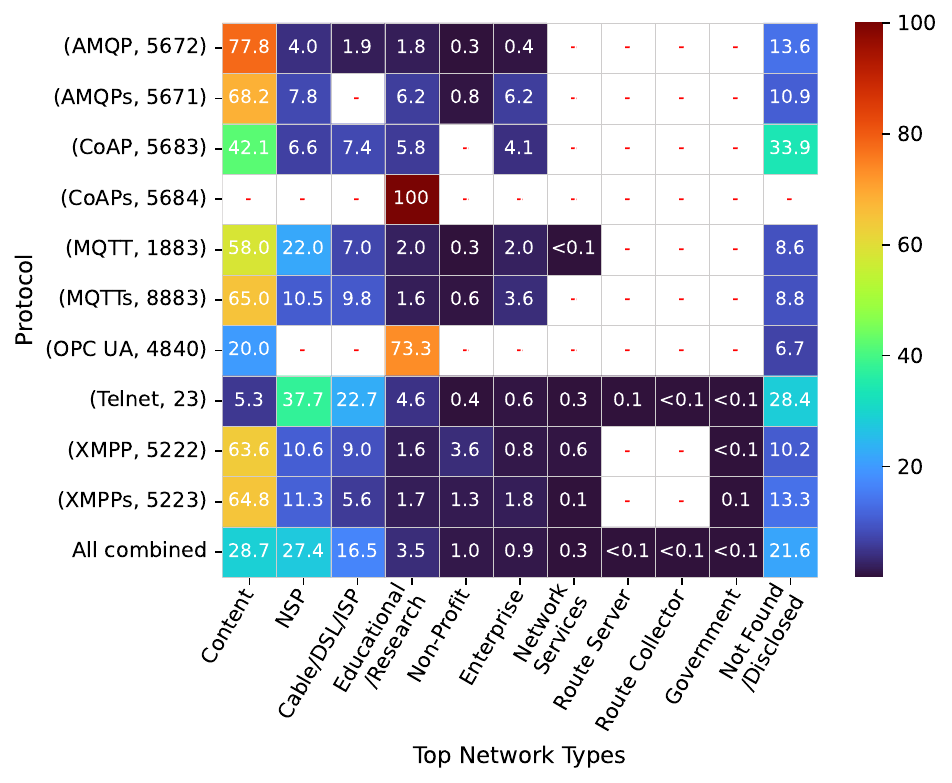}
    \caption{Heatmap \% IPs hosting our protocols in network types.}
    \label{f:networktype}
    \vspace{-0.3cm}

\end{figure}

\subsection{Characterizing the Networks}

In this section, we investigate the characteristics of the networks where our responding hosts are located. In particular, we are interested to identify the network types, such as whether they belong to content providers or eyeball networks. For this, we utilize the PeeringDB dataset~\cite{peeringdb}, a dataset that helps network operators to make decisions regarding peering requests at Internet Exchange Points (IXPs) or interconnection facilities. This dataset provides information about Autonomous Systems (AS) networks, including their network types and traffic volumes. We begin by mapping our IP addresses to AS Numbers (ASN) using the routeviews~\cite{routeviews} dataset. Then, we perform a lookup of the ASNs in the PeeringDB to identify the network types.

In~\Cref{f:networktype}, we present the percentage of IoT hosts for each protocol-port across all network types, again based on our latest measurement (2022-06-29). Firstly, for most protocols, the majority of IoT hosts are found within Content networks. This suggests that the observed IoT hosts likely function as IoT backend servers that support the IoT devices themselves~\cite{Saidi2022}. However, when examining Telnet hosts, we observe that they are predominantly located in Network Service Provider (NSP) and Internet Service Provider (ISP) categories, accounting for over \roughly{60\%} in both cases combined. Although we only observed a handful of CoAPs and OPC UA responding hosts, they were unexpectedly in educational/research networks. Additionally, for roughly 22\% of IoT hosts, no entry was found in PeeringDB (\char`\~17\%) or the network type was intentionally undisclosed(4.8\%). Overall, across all protocols (bottom row of the figure), the largest network category is Content providers \roughly{29\%}, closely followed by NSPs \roughly{27\%}.

\section{Security Analysis}

In this section we analyze different aspects of TLS security for the found IoT hosts:
We investigate the supported TLS versions, certificate issuers, expiration of certificates, self-signed certificates, and non-trusted certificates. We provide comparison of security analysis only for the protocols AMQPs, MQTTs, and XMPPs and not for OPC~UAs and CoAPs for which we do not discover more than one host. The only discovered OPC~UAs speaking host supports TLS 1.2 and uses an expired certificate. The CoAPs host we discover supports DTLS 1.2. 

\subsection{TLS Versions}

Since the first TLS standard was published in 1999 as TLS 1.0 \cite{rfc2246}, several updated versions of the protocol followed over the years \cite{rfc4346,rfc5246}, with the latest TLS 1.3 version \cite{rfc8446} being released in 2018.
Newer TLS versions come with improved security characteristics; therefore, the maximum supported TLS version can be used to assess the security posture of a host~\cite{paracha2021iotls}.
Hence, we analyze the maximum advertised TLS version from IoT hosts for the TLS-enabled versions of the AMQPs, MQTTs, and XMPPs protocols.

\begin{figure}[!t]
    \centering
    \includegraphics[width=.7\columnwidth]{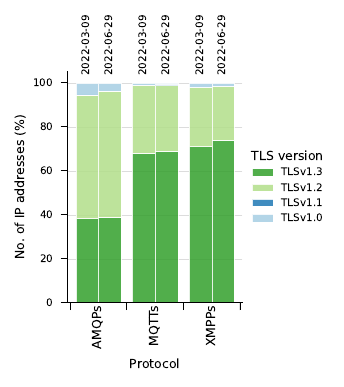}
    \caption{Comparison of the distribution of highest TLS version supported by secured TCP-based protocols AMQPs, MQTTs and XMPPs for two scans conducted three months apart.}
    \label{f:tls_version}
    \vspace{-0.3cm}
\end{figure}

In \Cref{f:tls_version}, we show the maximum advertised TLS version for AMQPs, MQTTs, and XMPPs for two measurements (March and June 2022).
We observe that for MQTTs and XMPPs, around 70\% of IPs offer the most recent TLS version 1.3.
In contrast, for AMQPs, we see a much lower prevalence of TLS 1.3 with just below 40\%.
The majority of the remaining share for all three protocols is allocated to TLS 1.2, with TLS 1.0 making up small single-digit percentages.
Moreover, we see a slight increase in TLS 1.3 support across all protocols in the three months period.

These relatively high deployment numbers of TLS 1.3 seem to be in line with Web server support \cite{lee2021tls,holz2020tracking,sosnowski2023dissectls}, but somewhat contrast the support of TLS 1.3 by IoT client devices, where we see a much lower prevalence of TLS 1.3 support~\cite{paracha2021iotls}.

\subsection{Certificate Issuers}

In total, we see 2239 hosts sending certificates for AMQPs, MQTTs, and XMPPs.
As has been studied before \cite{cangialosi2016measurement,gasser2018log}, the same certificate can be used on multiple hosts.
We find that a single certificate is sent by 247 hosts. 
Overall, the majority of certificates are unique, resulting in 1647 unique certificates.

Next, we analyze the issuers of TLS certificates sent by IoT IPs.
In \Cref{t:certificateIssuerTotalCombined}, we show the overall top 5 certificate issuers by the number of certificates.
We can see that overall, Let's Encrypt is the dominant issuer for IoT TLS certificates, spanning from 32\% to 77\%, depending on the protocol.
This is consistent with Let's Encrypt's dominance in the Web PKI \cite{farhan2023exploring}, where it is even more pronounced.

Interestingly, we find a relatively large number of certificates with ``Unknown'' issuers for AMQPs and XMPPs, \ie where the issuer organization field is empty.
We investigate this artifact in-depth and find that for AMQPs, the majority seem to be certificates created by the message broker software RabbitMQ\footnote{Issuer: \texttt{CN=TLSGenSelfSignedtRootCA, L=\$\$\$\$}, see this GitHub issue \cite{rabbitmq-github}.}.

In XMPPs, on the other hand, almost all of these are self-signed snake oil certificates~\cite{maghsoudlou2023characterizing}\footnote{Snake Oil certificates are self-signed certificates that are typically installed by default along with some packages or the operating system itself.} sent by addresses owned by Google and hinting at the lack of SNI used of our scans in the certificate's issuer fields\footnote{Issuer: \texttt{OU=No SNI provided\textbackslash; please fix your client., CN=invalid2.invalid}}.

\begin{table}[!t]
    \caption{Top 5 certificate issuers with share for AMQPs, MQTTs, and XMPPs.}
    \label{t:certificateIssuerTotalCombined}
    \centering
    \resizebox{\linewidth}{!}{
    \begin{tabular}{lrrrr}
        \toprule
        Issuer & Total & AMQPs & MQTTs & XMPPs\\
        \midrule
        Let's Encrypt & 61.1\% & 32.0\% & 77.4\% & 53.0\% \\
        Unknown & 18.8\% & 17.2\% & 2.9\% & 30.0\% \\
        Sectigo Limited & 5.3\% & 21.8\% & 2.1\% & 5.7\% \\
        DigiCert Inc &  1.7\% & 2.3\% & 0.7\% & 2.3\% \\
        ZeroSSL &  0.9\% & 0.9\% & 0.8\% & 0.8\%\\
        \bottomrule
    \end{tabular}
    }
    \vspace{-0.3cm}
\end{table}

\begin{table}[!t]
    \caption{Percentage of self-signed, expired, untrusted, and total number of certificates per IoT protocol.}
    \label{t:certificates}
    \centering
    \begin{tabular}{lrrrr}
        \toprule
        Protocol & Self-signed & Expired & Untrusted & Total\\
        \midrule
        AMQPs & 3.9\% & 25.0\% & 57.8\% & 128 \\
        MQTTs & 1.8\% & 23.7\% & 39.7\% & 858 \\
        XMPPs & 32.8\% & 13.0\% & 47.3\% & 1253 \\
        \bottomrule
    \end{tabular}
        \vspace{-0.3cm}

\end{table}

\subsection{Certificate Characteristics}

Finally, we investigate how many certificates belonging to IoT IPs are self-signed, expired, or untrusted, as shown in \Cref{t:certificates}.
We find that the share of self-signed certificates for AMQP and MQTT is below 4\%.
Interestingly, almost one-third of all XMPP certificates are self-signed.
We investigate this high number manually and find that they are again the self-signed snake oil certificates sent by addresses owned by Google.

\begin{figure}[!t]
    \centering
    \includegraphics[width=0.8\columnwidth]{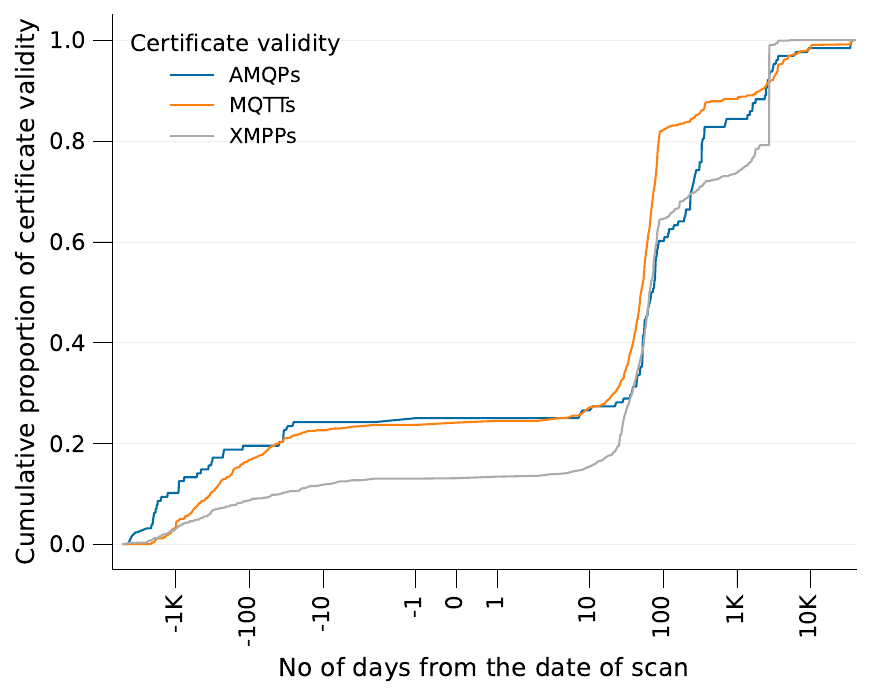}
    \caption{ECDF of validity in terms of certificates expiry used by secured TCP-based Protocols AMQPs, MQTTs and XMPPs.}
    \label{f:cert_expiry}
\end{figure}

Moreover, we analyze the expiry of certificates.
Again, we find similar percentages for AMQP and MQTT (around 25\%), whereas, for XMPP, the percentage is much lower at 13\%.
Manual investigation shows that this is due to a larger share of expired non-self-signed certificates for AMQP as well as MQTT (around 23\%) compared to XMPP (8.8\%).
We find that the top issuers for these expired certificates are Sectigo and Let's Encrypt for AMQP and MQTT, respectively.
We also analyze the number of days since a certificate has expired, as shown in \Cref{f:cert_expiry}.
We see that the majority of expired certificates have expired by more than 100 days, with a non-negligible share reaching more than three years (1k days).
We also see that the majority of non-expired certificates are valid for less than 90 days, hinting at the duration of certificates issued by popular free CAs such as Let's Encrypt, which we also see for the Web PKI \cite{farhan2023exploring}.

Next, using ZGrab2's runtime checks we analyze whether IoT TLS certificates are trusted by browsers.
We find that AMQPs has the largest share of untrusted certificates with almost 60\%, followed by XMPPs (47\%) and MQTTs (40\%).
When looking at these untrusted certificates in detail, we find that for AMQPs there are three reasons for untrusted certificates: unknown issuers as discussed in the previous section (43.2\%), signed by an untrusted CA (32.4\%), and expired certificate (24.3\%).
For MQTTs the most common reasons for untrusted certificates are expiry (53.4\%), unknown error (29.9\%), and signed by an untrusted CA (16.1\%).
Finally, for XMPPs the by far largest share is due to the snakeoil certificate sent by Google-owned IPs (70.0\%), followed by expired certificates (15.3\%) and signed by an untrusted CA (14.5\%).

\section{Related Work}

In this section we present a brief overview of related work and compare their results to ours.
Although there are numerous studies analyzing various aspects in the IoT ecosystem \cite{saidi2020haystack,guoIoT2020,pashamokhtari2021inferring,http-scan-iot,iotfinder,saidi2022badapple, Information-Exposure-IMC2019,Information-Exposure-IMC2019,Saidi2022}, most of them focus exclusively on the IPv4 Internet.

\parx{IoT deployment:}
In 2021, Srinivasa et al.~\cite{Shreyas2021} identified misconfigured IoT devices in the IPv4 Internet using active scan measurements. Their work primarily focused on scanning MQTT, AMQP, XMPP, CoAP, and Telnet on the standard unsecured ports.
Our work extends the list of protocols to include OPC~UA, a protocol that is extensively used in Industrial IoT \cite{opcua-ms}. Further, we scan the IPv6 Internet instead of IPv4 and plan to release all data and analysis scripts.
Comparing the results from our scan with the results published by Srinivasa et al., we find Telnet to be the most popular IoT-related protocol for both the IPv4 and IPv6 Internet. The second most popular protocol is XMPP for IPv6 as opposed to MQTT for IPv4. Considering the common protocols scanned on unsecured ports, the total number of IP addresses hosting IoT protocols for IPv4 is 13$M$ compared to 34.2$K$ for IPv6. However, it is worth to note that Srinivasa et al. considered port 2323 in addition to the standard port 23 for Telnet and the server port 5269 in addition to port 5222 for XMPP, which increases the chances of discovery for the Telnet and XMPP protocols.

In 2020, Dahlmanns et al.~\cite{opcua_vulnerability_scan} conducted an IPv4-wide scan to discover devices responding to OPC~UA probes.
They discovered a maximum of 2069 OPC~UA speaking IoT devices in a period of seven months from February to August 2020.
In our measurements spanning six months (January to June 2022), we discover a maximum of 19 OPC~UA speaking IoT devices in any particular scan.
We find the chances of a successful application-layer handshake on an open OPC~UA port to be higher for IPv6 (1.4\%) as opposed to IPv4 (0.5\%).

\parx{Vulnerability search engines:}
Shodan~\cite{shodan} and Censys~\cite{censys} are search engines that provide the data collected from Internet-wide scans by periodically performing active scans on TCP and UDP based protocols.
Historically, both search engines scanned only the IPv4 Internet.
More recently, they have slowly started releasing scan results for the IPv6 Internet for a few ports as well.
While Shodan does not publicly reveal their scanning pipeline, Censys initially used tools built on open source software such as the ZMap and ZGrab2~\cite{Zolotykh2021StudyOC,Bennett2021EmpiricalSA} to scan the IPv4 Internet. Neither Censys nor Shodan released their latest scanning tools for scanning the IPv6 Internet to be used by the research community.

Comparing the results from our scan with the results published by Shodan, we observe that Shodan has not discovered any OPC~UA speaking hosts in the IPv6 Internet.
Furthermore, for all the six protocols we have considered, Shodan has not managed to discover any devices on secured ports.
Moreover, the number of discovered IoT-speaking devices from our scans is 150 times higher (34$K$ vs. 225) than the data reported by Shodan, which indicates that Shodan has either an inefficient scanning methodology, it has been blocklisted by networks, or they do not use comprehensive target lists.

Comparing the results from our latest scan conducted on June 29, 2022, with the IPv4 data obtained from Censys for all the port-protocol combinations except for OPC UA~\cite{opcua_vulnerability_scan}, we find 100.6$K$ devices in IPv6 as opposed to Censys's 4.4$M$ in the IPv4 address space.
Furthermore, the total number of IP addresses hosting popular IoT protocols on their standard ports for IPv4 is 2.6$M$ compared to 36.4$K$ for IPv6.
Summing up these observations, around 59.7\% of IP addresses with open ports in IPv4 result in a successful application-layer handshake compared to 36.2\% in IPv6.
Hence, the chances of discovering protocols on their standard ports are lower in IPv6 compared to IPv4 for this hitlist.

\section{Conclusion}

\budget{0.25}

In this paper we performed an in-depth analysis of the IPv6 IoT ecosystem using active measurements.
We scanned 530M IPv6 addresses on six popular IoT-related protocols and found 36.4K IoT-speaking end-hosts in 156 countries.
In comparison with IPv4, we identified 380$\times$ fewer IoT-speaking end-hosts.
Our security analysis for TLS-enabled IoT protocols showed up to 57\% untrusted certificates, with up to 32\% being self-signed and 25\% being expired.
Finally, we plan to publish our results, tools, and web dashboard for further research.
\bibliographystyle{IEEEtranS}
\bibliography{paper}

\budget{0.75}

\end{document}